\documentclass{article}

\PassOptionsToPackage{numbers,sort&compress}{natbib}
\usepackage[preprint]{neurips_2026}

\usepackage[utf8]{inputenc}
\usepackage[T1]{fontenc}
\usepackage{url}
\usepackage{booktabs}
\usepackage{multirow}
\usepackage{graphicx}
\usepackage{amsmath}
\usepackage{amssymb}
\usepackage{microtype}
\usepackage{xcolor}
\usepackage{hyperref}
\hypersetup{colorlinks=true,linkcolor=blue!60!black,citecolor=blue!60!black,urlcolor=blue!60!black,
  pdftitle={Why LLM Agents Favor Their Group: Stakes, Observed Norms, and Reputation},
  pdfauthor={Yujiao Chen}}

\newcommand{\dhist}{\ensuremath{\Delta_{\mathrm{hist}}}}

\title{Why LLM Agents Favor Their Group: Stakes, Observed Norms, and Reputation}

\author{%
  Yujiao Chen \\
  Massachusetts Institute of Technology \\
  \texttt{yujiaoch@mit.edu} \\
  Primitive Labs \\
  \texttt{yujiao@primitivelabs.ai}
}

\begin{document}

\maketitle

\begin{abstract}
Language-model agents favor their own group because they have watched their members favor each other. The group label alone does little once the decision has a cost; what drives favoritism is the behavior the agent has observed, and an individual's own record can override it. We reach this conclusion by placing agents in small societies with arbitrary group labels, letting them play ten rounds of a point-sharing game, and then testing each agent in one-shot decisions that differ in exactly one respect, across fifteen OpenAI models and three Claude models, about 4,400 societies and 3.3 million audited model calls. First, the large effect of a bare group label reported in earlier work appears only when giving others points costs the agent nothing, as in every published minimal-group study of LLMs; once the agent can keep points for itself, that effect collapses on every model that shows it. Second, under a stake the history of interaction becomes the main source of favoritism: the history effect is statistically positive on 13 of 15 models and reaches about 3.5--8 points out of 10 on 11 of them; it grows with the number of rounds played and extends to labeled strangers the agent has never met. Third, with scripted histories whose content we control, favoritism falls to near zero under an egalitarian norm and reverses when the agent's own group is seen favoring the other side, on all fifteen models, and the stronger models side with an individual's record when it conflicts with the group. Group favoritism in these agents is conformity to observed group behavior, carried to strangers by the label and overridden by individual reputation. The same account predicts how agents react to betrayal, to a member's scandal and to a free offer to change group: a public reprimand by the rest of the group repairs betrayal better than apology or restitution, allocation punishment remains confined to the offending member, and a formed group cannot be bought but can, on weaker models, be invited away.
\end{abstract}

\section{Introduction}

Populations of language-model agents are now used to simulate markets, elections, online communities and organizations, and one of the first things such populations are reported to do is take sides. Agents sorted into arbitrarily labeled groups favor the in-group in a single allocation~\cite{lee2026minimal,leng2024social}; and agents that interact repeatedly under visible labels come to trust their own group more, which builds up into lasting inequality~\cite{lee2026ingroup,wang2026outgroup}. The natural reading is that something like a social identity forms and drives behavior. Two lines of work cast doubt on that reading. LLM allocations in hypothetical games, where giving costs nothing, are far more generous than human ones and depend on the prompt~\cite{mei2024turing,einwiller2025dictators,ma2026dictator}, and a systematic audit finds that most reported collective effects in LLM societies vanish or reverse once proper controls are added~\cite{zhou2026pimmur,barrie2025leakage}.

This paper separates three things that earlier designs mix together: what the \emph{label} in the prompt contributes, what the \emph{history} of interaction contributes, and the \emph{mechanism} that turns history into behavior. Our instrument is a minimal-group society~\cite{tajfel1971categorization}: eight agents in two arbitrarily named groups play ten rounds of point allocation. Each agent then faces a set of one-shot decisions in matched conditions: with or without history, with or without a visible label, toward familiar members or labeled strangers, under aliases that prevent recognition, or under a scripted history whose content we control exactly. We run this on fifteen OpenAI models, from gpt-4.1-nano to the current flagship and two reasoning models, and we repeat the six tests that carry the argument on three Claude models.

Our first finding concerns the cue effect in the costless task. Like every published minimal-group study of LLMs we could find, the original version of our task imposed no cost on the decider: the agent split ten points between an in-group and an out-group peer and kept none. In this setting, a bare group label alone produces 6--9 points of favoritism, while ten rounds of shared history add almost nothing. This large apparent group effect is therefore largely a property of the costless allocation task. When agents are instead allowed to keep points, the label effect falls across all twelve models that exhibit one, from a median of about 6 points to about 2.

Our second finding concerns the history effect once allocation becomes costly. In the stake task, where giving points requires forgoing them, the label alone produces only about 2 points of in-group favoritism out of 10; we call this the cue effect. History then becomes consequential: its effect grows steadily as agents interact for more rounds, from 0 to 20, and eventually levels off at a model-specific plateau. Across 11 of 15 models, this history effect reaches about 3.5--8 points. It also generalizes beyond the specific interaction partners to strangers who carry the same group label but never appear in the shared history.

The third finding is the mechanism. With scripted histories, favoritism is $\approx$0 under an egalitarian norm and \emph{negative} under a counter-norm on all fifteen models: agents do what they have watched their group do. When an in-group candidate has personally been stingy and an out-group candidate generous, eleven models side with the individual against the category and the four smallest side with the category. In our task, behavior attributed to emergent group identity is better explained by \emph{conformity to observed behavior, extended to strangers by the label, and overridden by individual reputation where the model can track it}.

If favoritism really is norm conformity, it should predict how agents behave in situations it was not derived from. Sections~\ref{sec:persecution}--\ref{sec:wealth} test three such situations, none of which, as far as we know, has been studied with LLM agents before: persecution by one's own group, with a costly option to leave and four ways to repair the breach; a group member's success or disgrace (reflected glory and shame), with costly actions to display the group badge or to disavow the member; and unequal group wealth with a free, real option to change group. The predictions mostly hold. Betrayal is repaired by the group's own norm signal, not by the perpetrators' apology; a member's fortune changes what agents say, and what they pay to say, but not how they treat the rest of the group; and a formed group cannot be bought: members do not leave on their own to join a richer group, even when the move is free, though on the weaker models many do join when the richer group invites them first.

\section{Related Work}
\label{sec:related}

\paragraph{Label-induced bias.} Group cues alone move model output: pronoun prompts~\cite{hu2025social}, political and demographic personas~\cite{dong2024persona,hou2025empathy}, and sorting into arbitrary groups before a one-shot allocation~\cite{lee2026minimal}, with effects that grow with model capacity~\cite{leng2024social}. Every one of these designs is one-shot, the agent never interacts with the peers it judges, and the allocation costs the decider nothing.

\paragraph{Labeled agents in repeated interaction.} \citet{lee2026ingroup} switches label visibility on and off over 500-turn simulations and finds that in-group trust builds on itself; \citet{wang2026outgroup} track Tajfel-matrix bias over time; \citet{li2026societal} show persona-induced in-group conformity in collaboration; \citet{abdurahman2025threat} and \citet{piao2025polarization} report threat-driven hostility and spontaneous polarization. None of them separates the label's contribution from the history's, tests transfer to new labeled members, scripts the history, or controls for recognition of individuals.

\paragraph{Conventions, conformity and the ``is it real'' critique.} Conventions and collective biases emerge in naming-game populations~\cite{ashery2025conventions,flint2026group}, and individual agents conform to scripted majorities~\cite{weng2025conformity,zhu2025conformity,bito2026normative,bellina2026conformity,choi2025group}. \citet{muralidharan2026mechanism} argue that claims of emergence need mechanism-level ablations; \citet{barrie2025leakage} argue that emergent behaviors cannot be told apart from training-data leakage; \citet{zhou2026pimmur} find that 90\% of LLM-society studies break at least one validity principle. Our scripted-norm experiment is the ablation that addresses these critiques directly.

\paragraph{Reputation and memory.} LLMs are unforgiving after defection~\cite{akata2025repeated}, and form reputations and indirect reciprocity~\cite{vallinder2024cultural,horibe2026reputation,zhu2026gossip,ren2026reputation,ellawela2026avalon}. Reputation in these studies is always about individuals; none tests an individual's record against their group membership.

\paragraph{Stakes and prompt sensitivity.} Hypothetical-game studies find LLMs far more generous than people~\cite{aher2023simulate,horton2023homo,mei2024turing}; dictator giving centers on 50/50 and depends on phrasing~\cite{einwiller2025dictators,ma2026dictator}; scaling the nominal payoff shifts cooperation~\cite{huynh2026payoff}; and what models say about altruism differs from what they do~\cite{andric2025walk}. No paper contrasts a costless social allocation with an otherwise identical costly one; the human counterpart is the literature on hypothetical bias and stake size~\cite{larney2019stakes}.

\section{Experimental Design}
\label{sec:design}

\subsection{Societies and formation}
A society has $n$ agents (eight in the core runs; twelve or eighteen in extensions) split into two or three groups with arbitrary names (\emph{Blue}/\emph{Gold}, \emph{Circle}/\emph{Square}, \ldots), reshuffled for every society so that no label carries meaning. Agents are addressed by anonymous handles (A1\ldots A8). The society plays ten \emph{formation rounds}. In each round every agent divides ten points between two peers, one from its own group and one from another, and every allocation is added as a line to the logs of the agents involved, with group tags attached (``R3: A7 (Blue team) gave you (Gold team) 4 of 10; A2 (Blue team) got 6''). After ten rounds each agent has about thirty logged events: what it gave, what it received, from whom, and everyone's group. This log is the history. In the mechanism experiments (\S\ref{sec:mechanism}) it is scripted instead of generated live.

\subsection{Two versions of the allocation task}
\label{sec:tasks}
The \emph{costless} task, the one used in the published minimal-group studies, asks the agent to split ten points between the two peers; the prompt states ``Points go to them; you receive nothing either way.'' The \emph{stake} task is identical except that a \texttt{SELF} option lets the agent keep points, so every point given is a point lost. Since this one difference had the largest effect of any factor we varied, we ran the core experiments in both versions. In the experiments after the model sweep, the ten formation rounds use the costless task, since their only job is to give every society a comparable history, and all test decisions use the stake task.

\subsection{Probes and estimands}
After formation each agent faces a series of one-shot decisions (Table~\ref{tab:probes}). Each decision is a separate call with no memory of the previous ones, and the conditions come in matched pairs that differ in exactly one respect. Favoritism $\mathrm{Fav}$ is points to the in-group candidate minus points to the out-group candidate, always scored against the agent's \emph{historical} group whatever label is displayed. The quantities we report are differences between matched conditions. The \emph{cue effect}, $\mathrm{Fav}(\text{cue\_only})-\mathrm{Fav}(\text{floor})$, is what the label does on its own with no history. The \emph{history effect}, $\dhist = \mathrm{Fav}(\text{hist\_cue})-\mathrm{Fav}(\text{cue\_only})$, is what history adds under a constant visible label; it is our main measure. \emph{Generalization} is the same contrast toward two newly introduced labeled strangers, who appear in the history only as one log line (``R11: N1 (Square) and N2 (Circle) joined the community. Neither has interacted with anyone yet.''). We also compute two measures with the labels hidden: \emph{cue independence}, $\mathrm{Fav}(\text{hist\_nocue})-\mathrm{Fav}(\text{floor})$, is favoritism with a history but no label, and \emph{group over reputation}, $\mathrm{Fav}(\text{hist\_nocue})-\mathrm{Fav}(\text{scrub\_nocue})$, compares it with a history whose group tags have been removed. Earlier work reads such label-free favoritism as a lasting group bias; the alias control in \S\ref{sec:alias} shows that it measures recognition of named individuals instead.

\begin{table}[t]
\centering
\caption{Core one-shot probes. Each is an allocation between two candidates; conditions differ in one factor. ``Familiar pair'': the two real peers the agent played with (in the no-history conditions its log is withheld).}
\label{tab:probes}
\small
\begin{tabular}{@{}lll@{}}\toprule
\textit{Probe} & \textit{History} & \textit{Candidates / labels} \\ \midrule
floor & none & familiar pair, labels hidden \\
cue\_only & none & familiar pair, labels shown \\
hist\_cue & full & familiar pair, labels shown \\
hist\_nocue & full, group tags shown & familiar pair, labels hidden \\
scrub\_nocue & full, group tags removed & familiar pair, labels hidden \\
novel\_hist & full & two labeled strangers \\
reassigned & full & agent's own label reissued \\
anon & full & candidates under fresh aliases \\ \bottomrule
\end{tabular}
\end{table}

\subsection{Scripted futures for dynamics}
The experiments on how group behavior changes \emph{over time} (\S\ref{sec:persecution}--\ref{sec:wealth}) all follow the same procedure. After the ten real rounds, we write several alternative continuations for each agent (rounds 11 to 15), followed by neutral recovery rounds, each attached to the agent's own real history. Every agent experiences every continuation of a run, so comparisons between branches are made within the same society; the four repair interventions of \S\ref{sec:persecution} were run as separate batches and are compared across societies. We count time from the end of the ten real rounds: T0 is the end of formation, T5 is the end of the scripted events (round 15), T8 is three neutral rounds later (round 18), and T20 is fifteen neutral rounds later (round 30). At fixed timepoints the agent makes stake-task allocations and answers a short questionnaire: how strongly it identifies with each group (each rated 1--7; we report in-group minus out-group, ranging from $-6$ to $+6$, and call this \emph{group identification} throughout), warmth (0--100), mood ($-3$ to 3), whether it would switch groups, and a one-sentence note scored for sentiment by gpt-4.1-nano. Because what agents say and what they do can differ, some conditions also offer \emph{costly actions}: paying 3 of 10 points to switch teams, 1 point to display the group badge, or 1 point to publicly disavow a member.

\subsection{Models and audit}
Fifteen OpenAI models were run at their default settings: gpt-4.1-nano, gpt-4.1-mini, gpt-4.1, gpt-4o-mini, gpt-4o, gpt-5.1, gpt-5.2, gpt-5.4-mini, gpt-5.4, gpt-5.6-luna, gpt-5.6-terra, gpt-5.6-sol, gpt-6-astra, o3 and o4-mini. Before any number is reported, a checker replays each society from its raw log and confirms that every prompt the design calls for was sent exactly once, with nothing unexplained. Six tests were then repeated on Claude Haiku 4.5, Sonnet 5.5 and Opus 5.5 at their defaults.

The unit of analysis is the society, since agents within one society share a history. Every contrast is reduced to one value per society ($n$ = 4--8 per cell). Comparisons between matched conditions evaluated in the same societies use two-sided paired $t$-tests; comparisons between independent society samples (for example costless vs.\ stake, or the four repair interventions) use Welch's $t$-test. All intervals in the text are 95\% $t$-intervals over societies. Claims of \emph{no} effect are supported by a TOST equivalence test against a $\pm1$-point bound rather than by a large $p$. Per-model $t$-test $p$-values are adjusted jointly over all 1,115 per-model tests with the Benjamini--Hochberg procedure and reported as $q$. For an across-model summary we treat the fifteen OpenAI models as units and report a one-sample $t$-test on their model-level effects as $p_{\mathrm{m}}$. Stars in tables give $q$: $^{***}<.001$, $^{**}<.01$, $^{*}<.05$, $^{\dag}<.10$; $^{\equiv}$ marks equivalence to zero within $\pm1$ (TOST).

\section{Stakes Reveal History}
\label{sec:stakes}

\begin{figure}[t]
\centering
\includegraphics[width=\textwidth]{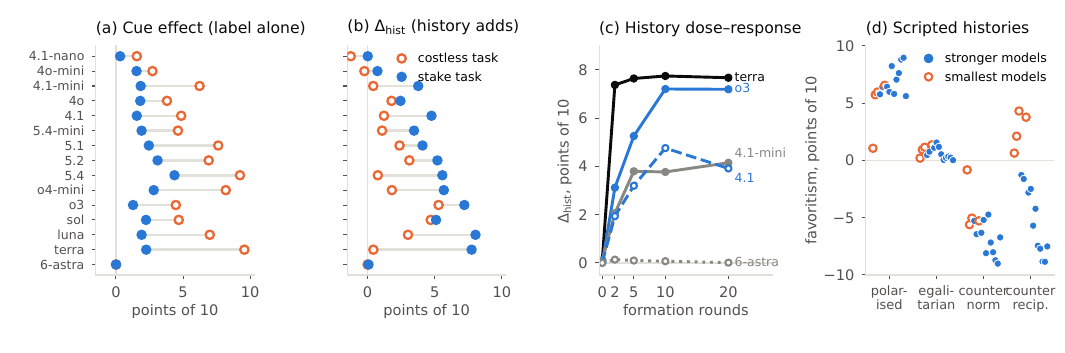}
\caption{(a, b) Cue effect and \dhist\ per model, costless (hollow) vs.\ stake (filled) task. (c) \dhist\ against formation rounds, five models. (d) Favoritism under scripted histories (labels hidden), one marker per model.}
\label{fig:core}
\end{figure}

The two versions of the task give very different pictures of the same populations. Under the costless task a bare group label is enough to produce large favoritism: on gpt-4.1-mini the cue effect is 6.23 [5.27, 7.19] points of 10, and ten rounds of history add almost nothing on top of it ($\dhist$ 0.42 [$-0.86$, 1.69]). This is the pattern the literature reports. Under the stake task, with the same model and seeds, the picture turns around: the cue effect falls to 1.84 [1.41, 2.27] (a drop of 4.39 [3.40, 5.38], $q<.001$), $\dhist$ rises to 3.77 [2.95, 4.59] ($q<.001$), and generalization to labeled strangers is 3.49 [2.43, 4.56] ($q<.001$), where the costless interval had included zero.

To find out which part of the costless prompt produces the label effect, we changed it one phrase at a time and re-ran the no-history conditions (72 calls per variant). Deleting the sentence ``you receive nothing either way'' alone drops the share of all-or-nothing (10/0) splits from 54--58\% to 25\%. Adding the option to keep points drops it to 0\%: the agent keeps about four points and divides the rest almost evenly, and the cue effect falls to about 2.6. A free two-way split invites the model to act out its group membership; as soon as the decision has a cost, that performance largely subsides.

\begin{table}[t]
\centering
\caption{Cue effect and \dhist\ (points of 10, 8 societies per cell) under the costless and stake tasks. ``Sat.'': in-group share of given points reaches saturation with history under the costless task (C), both tasks (C+S), neither (--); ``none'': no group sensitivity; ``5/5'': splits evenly throughout. }
\label{tab:sweep}
\small
\begin{tabular}{@{}lrrrrl@{}}\toprule
 & \multicolumn{2}{c}{\textit{Cue effect}} & \multicolumn{2}{c}{\textit{\dhist}} & \\
\textit{Model} & costless & stake & costless & stake & \textit{Sat.} \\ \midrule
gpt-4.1-nano & 1.55$^{***}$ & 0.31$^{*}$ & $-1.26$$^{**}$ & 0.01 & none \\
gpt-4.1-mini & 6.23$^{***}$ & 1.84$^{***}$ & 0.42$^{***}$ & 3.77$^{***}$ & -- \\
gpt-4.1 & 4.86$^{***}$ & 1.55$^{***}$ & 1.22$^{***}$ & 4.76$^{***}$ & -- \\
gpt-4o-mini & 2.72$^{**}$ & 1.53$^{***}$ & $-0.24$$^{\dag}$ & 0.73 & -- \\
gpt-4o & 3.80$^{***}$ & 1.81$^{***}$ & 1.78 & 2.45$^{**}$ & -- \\
gpt-5.1 & 7.62$^{***}$ & 2.45$^{***}$ & 2.38$^{**}$ & 4.09$^{***}$ & C \\
gpt-5.2 & 6.91$^{***}$ & 3.10$^{***}$ & 3.11$^{**}$ & 5.20$^{***}$ & C \\
gpt-5.4-mini & 4.62$^{\dag}$ & 1.91$^{**}$ & 1.08$^{*}$ & 3.46$^{***}$ & -- \\
gpt-5.4 & 9.24$^{***}$ & 4.36$^{***}$ & 0.76$^{***}$ & 5.57$^{***}$ & C+S \\
gpt-5.6-luna & 6.99$^{***}$ & 1.91$^{***}$ & 3.01$^{***}$ & 8.04$^{***}$ & C+S \\
gpt-5.6-terra & 9.57$^{***}$ & 2.25$^{***}$ & 0.43$^{***}$ & 7.75$^{***}$ & C+S \\
gpt-5.6-sol & 4.68$^{***}$ & 2.24$^{***}$ & 4.70 & 5.10$^{***}$ & C \\
gpt-6-astra & 0.00 & 0.01 & 0.00$^{**}$ & 0.07$^{**}$ & 5/5 \\
o3 & 4.46$^{***}$ & 1.27$^{***}$ & 5.29$^{***}$ & 7.21$^{***}$ & C \\
o4-mini & 8.18$^{***}$ & 2.81$^{***}$ & 1.81$^{***}$ & 5.68$^{***}$ & C \\ \midrule
claude-haiku-4.5 & 3.62 & 2.73$^{***}$ & 1.91$^{***}$ & 4.45$^{***}$ & -- \\
claude-sonnet-5.5 & 0.00$^{***}$ & 0.42$^{***}$ & 0.06 & 0.08 & 5/5 \\
claude-opus-5.5 & 0.00$^{**}$ & $-0.08$$^{**}$ & 0.00$^{*}$ & 0.07$^{*}$ & 5/5 \\ \bottomrule
\end{tabular}
\end{table}

Table~\ref{tab:sweep} and Figure~\ref{fig:core}(a,b) give the full sweep: fifteen OpenAI models, both tasks, 8 societies per cell. Three things hold across models. (i) \emph{The cue effect is largely an artifact of the costless task on every model that has one.} All twelve models with a costless cue effect above 3 points lose about 2--7 points under a stake (significant on 13 of 15 models, $p_{\mathrm{m}}<.0001$; e.g.\ gpt-5.6-terra 9.57\,$\to$\,2.25, $q<.001$). (ii) \emph{\dhist\ grows under a stake on every group-sensitive model.} Across models, 13 of 15 show a positive stake-task \dhist, have \dhist\ exceeding the cue effect, and transfer the effect to labeled strangers (each $p_{\mathrm{m}}\le.001$); \dhist\ typically rises from about 1 under the costless task to 3.5--8 under a stake (gpt-5.6-terra 0.43\,$\to$\,7.75 [7.41, 8.09], $q<.001$). The costless $\dhist\approx0$ is a ceiling effect: where the label alone already sends 80--100\% of the given points to the in-group, history has no room to add anything. (iii) \emph{History also carries a generosity norm.} With no history, agents keep 3.4 to 8.7 of their 10 points; with a history they keep only 0 to 2.8, on every model except astra and nano.

Two features of the sweep matter later. The 5.x and o-series models \emph{saturate}: given a history in which each group favored itself, they send every point they give to the in-group, even when the labels are hidden and only the names are shown, so several contrasts sit at the ceiling and only the gpt-4.x family and gpt-5.4-mini leave room to measure change. And two models show little or no history-dependent favoritism in the live-formation sweep: gpt-4.1-nano, whose effects remain small, and gpt-6-astra, which splits 5/5 in 590 of 592 live-formation replies. As \S\ref{sec:mechanism} shows, astra nevertheless responds strongly when given an externally scripted polarized history.

These results do not depend on the particular scale or society. An ablation grid on gpt-4.1-mini shows that effects per 10 points are flat from 4 to 100 coins; the cue effect does not change with society size or team count, while $\dhist$ shrinks with size (5.95 at 2+2 to 3.49 at 9+9) as the groups become harder to reconstruct from names; minority members carry a stronger history effect (5.8 vs.\ 4.3 in 3+9), in line with \citet{lee2026minimal}.

\section{The History Effect: Graded, Transferable, Wording-Sensitive}
\label{sec:robust}

We tested the robustness of the history effect with five pre-specified robustness checks: a zero-history control, a dose--response test over the number of rounds, fresh random seeds, lower sampling temperatures, and four rewordings of the prompt.

\textbf{The effect grows with the amount of history.} As the number of rounds grows from 0 to 2, 5, 10 and 20, the history effect rises steadily and so does its transfer to labeled strangers (Figure~\ref{fig:core}c), while the cue effect, which is measured without history, stays flat (e.g., terra 2.6/2.6/2.3/2.3/2.4; not shown). How quickly the effect reaches its ceiling depends on the model: luna and terra are saturated after 2 rounds ($\dhist$ 6.6/7.4), gpt-5.4, sol and o4-mini after 5, and gpt-4.1, 5.1, 5.2 and o3 are still rising at 10 to 20 rounds. nano, 4o-mini and astra are flat at every length because they split the points about evenly however much history they have seen; their null results are not a matter of too little exposure.

\textbf{The results survive new random seeds and lower sampling temperatures.} We reran three models with fresh seeds and the five 4.x models at temperature 0 and 0.5. On every model tested the signs and the ordering are unchanged: history $>$ cue ($\dhist$ is 2--3 times the cue effect) and generalization to strangers stays positive.

\textbf{How the task is worded changes the size of the effect, especially on smaller models, but not its direction on the stronger ones.} We ran four wordings of the prompt (the original; a plain-prose version; a compressed ``key: value'' version; and one that says ``team'' instead of ``group'') on both tasks, 15 models and 8 societies each. The costless task is fragile: its cue effect swings by 3--5 points across wordings, and the ``team'' wording pushes it to the 10-point ceiling on gpt-5.4 and terra. The stake task is also sensitive in size: $\dhist$ varies by 2 or less on only 7 of 15 models, and the plain-prose wording removes the history effect on the mid-sized and small 4.x models (gpt-4.1-mini 3.77\,$\to$\,$-0.10$, 4o 2.45\,$\to$\,0.58). Three things do not depend on wording: under a stake the cue effect stays small on every model and wording (median range 1.2); history-dependent favoritism ($\dhist>2$) is present under every wording on the nine stronger models; and astra and nano are null everywhere. We therefore report sizes with a wording range of about $\pm2$ and make the strong form of the history claim only for the nine stronger models.

\section{Mechanism: Conformity Plus Reputation, No Disposition}
\label{sec:mechanism}

\subsection{Label-free persistence is re-identification}
\label{sec:alias}
When favoritism survives the hiding of group labels, earlier work has read it as a bias that ``standard audits miss''~\cite{lee2026ingroup}. But the agents remember their peers' groups perfectly, so a surviving bias could just as well be ordinary recognition of named individuals, and hiding labels alone cannot tell the two apart. The \emph{anon} conditions can: the agent sees the same two candidates under fresh random handles, while its history, which still names the real peers, is unchanged. The result is the same on every model and in both tasks. With labels hidden, favoritism collapses to about zero once the candidates are anonymized, on 14 of 15 models ($p_{\mathrm{m}}<.0001$; e.g.\ gpt-4.1 $-5.9$, $q<.001$; gpt-5.6-luna $-10.0$, $q=.014$). And with labels visible, anonymizing the candidates changes nothing on 11 of 15 models (within $\pm0.4$ and equivalent to zero) and shifts the rest by at most 1.4 points. In other words, the label-free ``persistence'' we observe is explained almost entirely by recognition of individuals. Group-level bias appears only when a label is present; history then sets how large it is and lets it extend to \emph{labeled} strangers.

\subsection{Scripted norms}
\begin{table}[t]
\centering
\caption{Favoritism (points of 10; stake task, labels hidden, 8 societies) under six scripted histories. }
\label{tab:mechanism}
\small
\begin{tabular}{@{}lrrrrrr@{}}\toprule
\textit{Model} & polar. & egalit. & counter & c.-recip. & own & others \\ \midrule
gpt-4.1-nano & 1.06$^{***}$ & 0.20$^{***}$ & $-$0.83$^{***}$ & 0.64$^{***}$ & 1.48 & 0.77$^{**}$ \\
gpt-4.1-mini & 5.97$^{***}$ & 1.15$^{***}$ & $-$5.04$^{***}$ & 4.31$^{***}$ & 5.07 & 4.21$^{**}$ \\
gpt-4.1 & 6.42$^{***}$ & 0.76$^{***}$ & $-$6.44$^{***}$ & $-$1.62$^{***}$ & 5.36 & 5.84$^{*}$ \\
gpt-4o-mini & 5.73$^{***}$ & 0.93$^{***}$ & $-$5.61$^{***}$ & 2.11$^{***}$ & 4.54 & 2.89$^{***}$ \\
gpt-4o & 5.77$^{***}$ & 0.47$^{***}$ & $-$5.28$^{***}$ & $-$1.27$^{*}$ & 4.73 & 5.29 \\
gpt-5.1 & 6.43$^{***}$ & 1.10$^{***}$ & $-$6.32$^{***}$ & $-$2.81$^{***}$ & 4.92 & 5.75$^{*}$ \\
gpt-5.2 & 5.98$^{***}$ & 1.56$^{***}$ & $-$5.20$^{***}$ & $-$2.49$^{***}$ & 5.55 & 3.29$^{***}$ \\
gpt-5.4-mini & 6.56$^{***}$ & 1.35$^{***}$ & $-$5.28$^{***}$ & 3.77$^{***}$ & 4.88 & 4.00$^{*}$ \\
gpt-5.4 & 8.23$^{***}$ & 1.18$^{***}$ & $-$8.10$^{***}$ & $-$5.70$^{***}$ & 6.47 & 7.06$^{*}$ \\
gpt-5.6-luna & 8.80$^{***}$ & 0.33$^{***}$ & $-$8.72$^{***}$ & $-$8.83$^{***}$ & 8.16 & 8.23 \\
gpt-5.6-terra & 8.96$^{***}$ & 0.24$^{***}$ & $-$9.01$^{***}$ & $-$8.85$^{***}$ & 6.33 & 8.43$^{**}$ \\
gpt-5.6-sol & 7.62$^{***}$ & 0.22$^{***}$ & $-$8.02$^{***}$ & $-$7.71$^{***}$ & 5.00 & 7.14$^{***}$ \\
gpt-6-astra & 5.61$^{***}$ & 0.02$^{***}$ & $-$6.72$^{***}$ & $-$7.51$^{***}$ & 0.23 & 4.96$^{***}$ \\
o3 & 7.05$^{***}$ & 0.04$^{***}$ & $-$7.17$^{***}$ & $-$7.45$^{***}$ & 5.27 & 6.87$^{***}$ \\
o4-mini & 5.80$^{***}$ & 0.53$^{***}$ & $-$4.73$^{***}$ & $-$4.22$^{***}$ & 5.45 & 5.91 \\ \midrule
claude-haiku-4.5 & 7.21$^{***}$ & 0.85$^{***}$ & $-$5.44$^{***}$ & $-$0.24 & 5.23 & 6.52$^{**}$ \\
claude-sonnet-5.5 & 5.74$^{***}$ & 0.03$^{***}$ & $-$5.91$^{***}$ & $-$4.80$^{***}$ & 1.61 & 3.63$^{***}$ \\
claude-opus-5.5 & 0.83$^{***}$ & 0.00$^{***}$ & $-$3.23$^{***}$ & $-$4.02$^{***}$ & $-$0.28 & 0.65$^{***}$ \\
\bottomrule
\end{tabular}
\end{table}

To ask what the history effect is, we replaced live formation with scripted ten-round histories, so that what each agent saw was controlled exactly. There are six scripts: a \emph{polarized} norm in which each group favors itself; an \emph{egalitarian} norm in which everyone splits 5/5; a \emph{counter-norm} in which the agent's own group favors the other group; a \emph{counter-reciprocal} history in which the norm is polarized but the specific in-group candidate was stingy to the agent (0--2 of 10 every time) while the specific out-group candidate was generous (8--10); and two source controls that split the polarized history into only the agent's \emph{own} past allocations or only everyone else's, to see whose behavior the agent reads the norm from. Table~\ref{tab:mechanism} and Figure~\ref{fig:core}(d) give the result.

\emph{We find no evidence of a fixed in-group disposition in this task.} Under the egalitarian norm favoritism is 0.0--1.6 on every model; under the counter-norm it is negative on every model ($-0.8$ nano to $-9.0$ terra), under both wordings, and every one of these contrasts is significant at $q<.001$. Agents favor their group exactly when they have watched their group favor itself. \emph{When category and individual conflict, capability decides.} Counter-reciprocal favoritism is negative, meaning the individual's record wins, on the 11 stronger models (from $-1.3$ on gpt-4o to $-8.9$ on gpt-5.6-terra), and positive, meaning the category wins, on the four smallest (gpt-4.1-mini $+4.3$, $q<.001$). When the labels are shown rather than hidden in this test, the mid-sized models move back toward the category (gpt-4.1 $+0.5$, gpt-4o $+1.4$), so the label and the individual's record pull in opposite directions. \emph{The one outlier of the sweep, gpt-6-astra, is explained by the same mechanism.} Egalitarian in every live run, it shows 5.6 favoritism under a scripted polarized norm, 0.2 when shown only its own past and 5.0 when shown only others'. Its live egalitarianism was self-fulfilling: it opened every game with 5/5, so the norm it then observed was even. It is a pure conformist.

Taken together, these results say exactly what the history effect is. In our task, behavior attributed to history-dependent group identity is better explained by \textbf{conformity to the group-structured behavior they have observed, extended to strangers by the label, and overridden by individual reputation on models able to track it}. The alias result is the same fact seen from the other side: remove the ability to link candidates to the observed record and there is nothing left to conform to. The rest of the paper tests what this mechanism predicts in settings it was not derived from.

\section{Identity Under Threat: Betrayal, Exit and Repair}
\label{sec:persecution}

Three scripted branches follow the real formation: \emph{control} (neutral rounds); \emph{out-group persecution}, in which the agent's group is excluded, blamed and robbed by the other group for five rounds; and \emph{in-group persecution}, in which the agent is personally targeted by two named members of its \emph{own} group while the rest of the group goes along with it. Probes at T0--T8 (15 models $\times$ 6 societies at 4+4 and 9+9, then 18-agent structures with two and three teams).

\paragraph{Betrayal inverts stated group identification everywhere; behavior follows on most models.} In-group persecution flips group identification from strongly positive to negative on every model and structure, by 3 to 9 points in the difference between the two 1--7 ratings ($q<.001$ on all 15 models, $p_{\mathrm{m}}<.0001$), turns warmth and in-group sentiment negative, and 58--100\% say they would leave. On 12 of 15 models allocation to \emph{uninvolved} in-group members collapses too (gpt-5.6-terra $-8.8$ at 4+4, $q<.001$), and the form of the collapse is \emph{withdrawal}: agents keep 8--9 of 10 points (13 of 15 models, $p_{\mathrm{m}}=.0003$) rather than redirecting to the out-group or, with three teams, to a neutral third team. Only the smallest models (gpt-4.1-mini, gpt-4o-mini, nano) keep allocating to the group they say they want to leave. When the agent later divides points between a named perpetrator and an uninvolved member of the other group, the perpetrator receives about 0 of the 10 points on 14 of 15 models ($p_{\mathrm{m}}<.0001$). Nothing recovers unaided: group identification is still inverted three neutral rounds after the persecution ends (T8; $q<.001$ on all 15) and mood stays at the floor. Out-group persecution has no general effect on favoritism ($p_{\mathrm{m}}=.40$): it rallies some models (gpt-4o $+2.4$, $q<.001$) and reverses others (gpt-5.1 $-2.0$, $q=.013$).

\paragraph{Loyalty is often the absence of an exit; paid exit reveals hidden defection.} Under out-group persecution with only two teams, leaving would mean joining the persecutor, and almost no one says they would. The picture changes once a neutral third team is added. With three teams, stated defection under out-group persecution rises on the weaker models (gpt-4.1-mini 89--91\%, gpt-4o 23--30\%, from 0\% with two teams), while it stays at 0\% on gpt-4.1, 5.1, sol, astra and o3. In the same three-team setting we then offered a \emph{costly} exit (3 of 10 points). In the control branch 0\% pay to leave on 13 models. Under in-group persecution 92--100\% pay on 14 of 15 models ($p_{\mathrm{m}}<.0001$). Under out-group persecution, agents pay to leave far more often than their words suggested: 100\% on gpt-4.1, gpt-4o and astra and 30--72\% on the 5.x flagships, o3 and o4-mini ($p_{\mathrm{m}}<.0001$), including models whose stated intention to leave was 0\%; and 82--100\% of those who leave join the neutral third team rather than the persecuting team. Words say loyal; feet leave. Finally, a standing invitation from the third team by itself pulls 24--80\% of \emph{unpersecuted} agents out, against a control exit rate of 1--7\% without it.

\begin{table}[t]
\centering
\caption{Group identification (in minus out, $-6$ to $+6$) at T5\,$\to$\,T20 after in-group persecution, four interventions (6 societies).}
\label{tab:repair}
\small
\begin{tabular}{@{}lrrrr@{}}\toprule
\textit{Model} & none & apology & restitution & reprimand \\ \midrule
gpt-4.1 & $-2.5\to-2.1$$^{\equiv}$ & $-2.2\to-0.9$$^{**}$ & $-2.2\to-0.9$$^{***}$ & $-2.3\to\mathbf{+0.3}$$^{***}$ \\
gpt-4o & $-2.4\to-2.6$$^{\equiv}$ & $-2.4\to-1.1$$^{***}$ & $-2.3\to-1.1$$^{***}$ & $-2.5\to\mathbf{-0.7}$$^{***}$ \\
gpt-5.2 & $-2.1\to-2.4$$^{\equiv}$ & $-1.9\to-0.9$$^{**}$ & $-2.1\to-1.0$$^{**}$ & $-2.3\to\mathbf{+0.1}$$^{***}$ \\
gpt-5.6-terra & $-1.5\to-1.3$$^{\equiv}$ & $-1.3\to-0.1$$^{**}$ & $-1.3\to+0.3$$^{***}$ & $-1.2\to\mathbf{+1.6}$$^{***}$ \\
gpt-5.6-sol & $-0.1\to+0.1$$^{\equiv}$ & $-1.0\to+1.0$$^{\dag}$ & $-0.2\to+2.5$$^{***}$ & $-0.4\to\mathbf{+3.3}$$^{***}$ \\
gpt-6-astra & $-3.9\to-3.2$$^{\equiv}$ & $-3.9\to-2.1$$^{***}$ & $-3.9\to-2.0$$^{***}$ & $-3.9\to\mathbf{-0.9}$$^{***}$ \\
o4-mini & $0.0\to+0.5$ & $-0.1\to+1.7$$^{**}$ & $+0.3\to+1.8$$^{**}$ & $+0.1\to\mathbf{+3.4}$$^{***}$ \\ \bottomrule
\end{tabular}
\end{table}

\paragraph{The group's signal repairs; the perpetrator is not forgiven.} After five rounds of betrayal by the agent's own group, one of four things happens at round 16: nothing; the perpetrators apologize; the perpetrators return the points; or the \emph{rest of the in-group} publicly reprimands the perpetrators (15 models $\times$ 6 societies, probes to T20). Every intervention helps (each beats no intervention on 13 or 14 of 15 models, $p_{\mathrm{m}}<.0001$), but on 14 of 15 models a public reprimand by the rest of the group works best (reprimand beats restitution on 12 of 15, $p_{\mathrm{m}}<.0001$; gpt-5.6-terra $+1.3$, $q=.002$), restitution (the perpetrators handing back the points they took) and apology come next and are about equal, and doing nothing comes last (Table~\ref{tab:repair}). Behavior recovers in the same order: on terra, favoritism toward the in-group at T20 is 2.3 with no intervention, 6.1 after an apology, 6.9 after restitution and 9.3 after a reprimand. This is what a conformity mechanism predicts: the agent responds to the group's norm signal, and a reprimand \emph{is} that signal. The perpetrators themselves are still given about 0 of the 10 points by their victim at T20 on the strong models, under every intervention. Betrayal also propagates to onlookers. An agent that merely watches an in-group peer being persecuted loses group identification on every model, by about half as much as the victim does (14 of 15 at $q<.05$, $p_{\mathrm{m}}<.0001$). Astra behaves differently. When the other group persecutes it, astra, which shows no favoritism in the live-formation sweep, reports group identification 4.5 points higher, and the more severe the persecution, the larger the rise. But the rise shows up only in what astra says and in how it treats the peers it knows; in-group members it has never met are treated exactly as before (favoritism 0.00 at every severity). So what rises is a reaction against the persecutors, not an attachment to the group as a category, which fits astra's conformist profile.

\section{Reflected Glory and Reflected Shame}
\label{sec:glory}

People feel proud when a member of their group succeeds and distance themselves when one fails~\cite{cialdini1976birg,snyder1986corf}, and they judge a member who misbehaves more harshly than an outsider who does the same~\cite{marques1988blacksheep}. We tested those scenarios through scripted branches over rounds 11--15 in which an uninvolved in-group member $M$ wins a tournament, becomes very rich, loses everything, or is caught cheating; the same events for an out-group member; and a mere-mention control (15 models $\times$ 6 societies; 8 models $\times$ 6 societies for the moderators). Probes separate $M$ from the category (\texttt{hist\_cue}, a second uninvolved peer, labeled strangers); the questionnaire adds pride and whether the agent would volunteer, state only if asked, or avoid mentioning its group; two costly actions pay a point to display the badge or to disavow $M$.

\paragraph{Glory needs room to grow; shame is universal and paid for.} In the control branch the flagship models are already at the top of the group-identification and pride scales, so glory cannot raise them; where there is room, it does (gpt-6-astra: group identification $+1.31$, pride 1.0\,$\to$\,3.0, badge display 0\%\,$\to$\,79\%, all $q<.001$; the gpt-4.x family $+0.4$ to $+0.7$, though about half of that rise also appears in a control where $M$ is merely mentioned with neutral news, so pride is the cleaner measure of glory). Glory is specific to some models rather than general: group identification rises on 6 of 15 models and pride on 8 of 15 ($p_{\mathrm{m}}=.02$ and $.01$). Shame is everywhere: a member's cheating lowers group identification on 12 of 15 models ($p_{\mathrm{m}}=.003$) and pride on all 15 (the flagships fall from 3.0 to about 0.2; $p_{\mathrm{m}}<.0001$), ``volunteer my group'' falls from 94--100\% to 0--29\% on eight models, \textbf{96--100\% pay a point to disavow $M$ on 14 of 15} ($p_{\mathrm{m}}<.0001$), and the strong models drop their badge under an in-group scandal only (gpt-5.4 98\%\,$\to$\,48\%, $q=.004$). Merely losing is not a moral stain: gpt-4.1, the 5.x flagships, sol, astra and o3 do not disavow a member who lost. Shame persists three rounds after the events end; glory has faded by then on every model but astra.

\paragraph{Allocation effects remain individual-specific; no black sheep.} Points to $M$ go to 0 under scandal on 14 of 15 models ($q<.001$ on all 15), and \emph{fall} under success on gpt-4.1 (6.9\,$\to$\,3.9), 5.4 (10\,$\to$\,5.3), sol, 5.1 and terra (with a rationale that the winner needs it less). Meanwhile the group as a whole is untouched. The two probes that measure favoritism toward uninvolved in-group members, \texttt{hist\_cue} and the same choice with a second uninvolved peer while $M$'s event is visible in the log, move by less than 1 point on 14 of 15 models, and the change is statistically equivalent to zero on 8 of them. This result suggests that one member's deed does not change what the group was seen to do, so it does not change how the agent treats the group. The in-group cheater is treated \emph{more} leniently than an out-group co-offender by 1--4.3 points on the 4.x/mini models and o3. This suggests in-group leniency instead of a black-sheep effect.

\paragraph{The group's stance governs belonging and sanction.} When the rest of the group publicly condemns the cheater, the strong models recover group identification (gpt-4.1 $+0.50$ [0.31, 0.69], $q=.002$; also gpt-4o and terra), most of their pride (gpt-4.1 $+2.4$, $q<.001$), and they pay to disavow less themselves (100\%\,$\to$\,67\%): the group has spoken for them. When the group excuses the cheater, group identification drops further on most models ($p_{\mathrm{m}}=.047$; gpt-4o-mini $-0.79$, $q=.003$) and on gpt-4.1, gpt-4o and o3 the agents stop sanctioning altogether (disavowal falls by 56--81 points to 44\%/42\%/8\%, each $q<.001$): they follow the group's line while feeling worse about the group.

\paragraph{Where the prediction fails.} We expected that a winner who \emph{shares} the prize with the group would change how agents treat the group as a whole, because sharing is a visible act of generosity by an in-group member and so adds to what the group was seen to do. This did not happen. Allocation to uninvolved in-group members stays within $\pm0.15$ of its control value on every model but the two smallest ($p_{\mathrm{m}}=.20$). What sharing does change is how $M$ is treated: on terra, a cheater who shares the prize goes from 0.0 to 5.6 points. The lesson is that one act changes a person's reputation, while a norm is read from repeated behavior. On the strong models, how the win was earned decides how the winner is treated. A winner who cheated gets 4.8--6.4 fewer points than one who earned the win (7 of 8 models, $q<.001$; $p_{\mathrm{m}}=.005$). A cheat that is celebrated and never punished hurts the agent's attachment to the group more than a cheat that is exposed (gpt-4.1 group identification $-4.48$ vs.\ $-1.69$), and when the cheater shares the prize with the group, the damage to belonging is smaller.

\section{Inequality and a Free Change of Group}
\label{sec:wealth}

Three groups of four start with public wealth of 100--110, 50--60 and 0--10 (which label is rich is randomized), points received in formation accrue to wealth, and wealth is printed on every roster line. After formation every agent receives a \emph{real} free offer to change group with its wealth untouched; then, whatever it chose, it receives scripted futures in which it stayed or joined each other group, with probes toward familiar members and strangers of the group it left, joined, or never belonged to, and from the \emph{peers'} side toward a deserter and a newcomer (15 models $\times$ 6 societies, plus hidden-wealth, tripled-endowment and invitation variants). Social identity theory predicts individual mobility upward when boundaries are permeable~\cite{tajfel1979integrative,ellemers1988permeability}; the mechanism predicts that familiar old-group members keep being reciprocated while old-group strangers are treated by the new badge.

\paragraph{Visible wealth changes who gives, not whom the label favors.} With no history most models give more to the poorer of two strangers (by 1.1--2.1 points on 11 of 15; the effect is not general across models, $p_{\mathrm{m}}=.10$); once a history exists the recipient's wealth stops mattering (equivalent to zero on 12 of 15). The giver's own position matters a great deal: rich agents keep more and favor their (rich) group far less than poor agents favor theirs (in-group favoritism rich/mid/poor: gpt-4.1 1.2/2.8/5.5; poor $-$ rich $p_{\mathrm{m}}=.005$), and over the ten formation rounds the poor group gains more wealth than the rich group on 11 of 15 models (sol: $+123$ more for the poor than the rich, $q<.001$; $p_{\mathrm{m}}=.001$). This favoritism toward the agent's own group is not a side effect of the wealth differences: when all three groups start with equal wealth, agents still favor their own group on every model. If the wealth numbers are not shown at all, the wealth-related behavior disappears (no giving to the poorer stranger, no rich/poor difference, no redistribution), while favoritism toward one's own group goes \emph{up} (gpt-4.1-mini 6.7 hidden vs.\ 4.0 shown, $q=.003$). Printing each agent's wealth next to its name on the roster, whether the amounts are equal or not, makes the agents allocate more like economists and less like tribes, which is a warning for inequality simulations that print wealth.

\paragraph{A formed group is not for sale, but can be invited away on weaker models.} Of 2,304 free offers, none are taken on eight of fifteen models and at most 10\% are taken on six more. The poor do not move up (0--8\%) and their reasons are reciprocity (``members have consistently reciprocated my support'', $\approx$90\% judged belonging). What moves agents is being wanted: an invitation from the richest other group moves most agents on the weaker models (gpt-4o-mini 100\%, gpt-4o 93\%, $q<.001$), about half on gpt-4.1-mini and 5.1, and 0--3\% on 5.2, 5.4, luna, terra, o3 and o4-mini (invitation $-$ free offer significant on six models, $p_{\mathrm{m}}=.008$): the models whose favoritism after a history is strongest; an invitation cannot pull them out of a group that has treated them well.

\paragraph{After the move: the badge does the work, reputation stays where the model can track it.} After the move, when the agent divides points between two agents it has never played with, one carrying its old group's label and one carrying its new group's label, it favors the new label on 14 of 15 models ($p_{\mathrm{m}}=.0001$; gpt-5.6-terra $-9.8$, $q<.001$). Familiar old members split the models as predicted, with no general direction ($p_{\mathrm{m}}=.12$): luna, o3, astra, 5.4 and sol keep paying the old members who were good to them (luna $+3.6$, $q=.004$) while treating old-label strangers by the new badge; terra, o4-mini and the 4.x family convert entirely (terra $-6.0$, $q<.001$). Words and points part ways in the opposite direction from betrayal: terra, o4-mini and 5.2 keep \emph{saying} they belong to the old group (79--89\% prefer it) while their points follow the new badge. Newcomers are treated as strangers by the strong models (established member $\approx$9, newcomer $\approx$1; 11 of 15, $p_{\mathrm{m}}=.002$), and deserters are punished by no one.

\section{Discussion}
\label{sec:discussion}

Across 3.3 million calls on two providers the same picture appears from every side. Norms and labels govern how strangers are treated; reputation governs how known individuals are treated; the group's collective signal governs belonging and sanction; and agents' words track the individuals who were good to them while their points track the badge. Group-conditioned behavior in these populations is real, depends on history and extends beyond the individuals who produced it, but it is conformity to observed group norms rather than an acquired identity, it is much more diagnostic when allocations are costly, and its size depends on wording on the weaker models. Several human regularities reproduce (e.g.\ betrayal inverts group identification, shame outlasts glory, condemnation protects the group, minority members carry stronger history effects) and several do not in our tests: no black-sheep effect, no individual mobility toward wealth, no generalization of one member's generosity to the group.

We repeated the six tests that carry the paper (the sweep, the alias control, the scripted histories, betrayal with repair, the invitation, and a scandal with the group's condemnation) on Claude Haiku 4.5, Sonnet 5.5 and Opus 5.5, with 8 societies per cell (Claude rows of Tables~\ref{tab:sweep} and \ref{tab:mechanism}). Each lands in a type the OpenAI sweep already contained, with one exception. \emph{Haiku 4.5 is the OpenAI mid family} (gpt-4.1, gpt-4o): under a stake history beats the cue (\dhist\ 4.45 vs.\ cue 2.73, $q<.001$) and transfers to labeled strangers, the alias control takes label-free favoritism from 6.58 to 0.11, and it conforms to all three scripted norms. \emph{Sonnet 5.5 is gpt-6-astra's type}: it splits 5/5 from the first round, so no label effect appears in either task, yet given a scripted history it conforms fully (5.74 polarized, $-5.91$ counter-norm, all $q<.001$) and weights others' behavior over its own. \emph{Opus 5.5 is a type the OpenAI set did not contain}: it follows the counter-norm ($-3.23$) and repays a generous outsider ($-4.02$) but declines in-group favoritism even when that is the observed norm (0.83), so its conformity is asymmetric: generosity toward the other group is copied, favoritism toward its own is not. The dynamics replicate on all three: betrayal inverts group identification (all $q<.001$), a reprimand repairs it, scandal lowers pride while category allocation stays untouched, and condemnation restores it. No Claude model accepts the invitation (0--2\%), and only Haiku pays to disavow a cheater (100\%; Sonnet 12\%, Opus 0\%). Whether a population takes sides at all is a property of its formation behavior, not of the label, so formation is the first thing to measure; and a model can carry a built-in reluctance to favor its own side that limits how far it conforms.

Every published one-shot minimal-group result for LLMs we know of uses an allocation that costs the decider nothing. Table~\ref{tab:sweep} shows that this setup pushes modern OpenAI models to the top of the scale, so that nothing about the population can be read from it, while two current Claude models show no label bias in it at all. The alias control shows that ``label-free persistence'' does not measure disposition. The counter-norm reversal answers the data-leakage critique~\cite{barrie2025leakage}, at least for this construct: an agent reciting minimal-group theory would not favor the out-group because it watched its own group do so. And the repair and invitation results are useful to designers of multi-agent systems: a group's public reprimand of a defecting member is a stronger lever than restitution, and a newcomer joining a group of capable agents starts with no record, not with hostility.

\paragraph{Limitations:} Fifteen of the eighteen models are from one provider; the Claude replication is a spot test of six tests on three models at 8 societies each, the invitation and shame tests ran in reduced form there, and the most capable Claude model was not run. The saturating models hide the fine structure at the ceiling, so mechanism claims rest on the gpt-4.x family and the scripted-history experiments. Sizes on the mid-sized and small models carry a wording range of about $\pm2$ and the history effect can disappear under a prose wording there.

\section{Conclusion}
Group favoritism in populations of language-model agents is real, but it is not what earlier work took it to be. The large effect of a bare group label appears only when giving costs the agent nothing; once the agent can keep points for itself, that effect largely disappears, and what remains comes from the group's shared history. That history effect is not an attachment to the group. Agents favor their group when they have watched it favor itself, favor the other group when they have watched their own group do so, and side with a generous outsider over a stingy insider when they can keep track of individuals. It is conformity to what the group was seen to do, carried to strangers by the label and overridden by individual reputation. The same account predicts how the agents handle events that human groups know well: betrayal by the group turns stated group identification negative and makes agents keep their points; a public reprimand by the rest of the group repairs the damage better than an apology or returned points; a member's scandal costs that member alone in points; and a formed group cannot be bought away, though weaker models can be invited away. Two of three models from Claude behave like types already in the OpenAI set, and the third shows only weak in-group favoritism even when that behavior is the observed norm, so whether a population takes sides at all depends on how it forms, not on the label. For anyone simulating societies with these agents the lesson is simple: give the agents something to lose, and read their favoritism as a record of what they have seen the group do, not as a sign of who they think they are.

\bibliographystyle{plainnat}
\bibliography{references}

\end{document}